\documentclass[journal=nalefd,manuscript=article,layout=traditional]{achemso}
\usepackage[version=3]{mhchem}
\usepackage{xcolor}

\title{Large-Gap Quantum Spin Hall Insulator in single layer bismuth monobromide $\bf{Bi}_4\bf{Br}_4$ }

\author{Jin-Jian Zhou}
\affiliation[Beijing Institute of Technology]
{School of Physics, Beijing Institute of Technology, Beijing 100081, China}
\alsoaffiliation[Institute of Physics, Chinese Academy of Sciences and Beijing
National Laboratory for Condensed Matter Physics]
{Institute of Physics, Chinese Academy of Sciences and Beijing
National Laboratory for Condensed Matter Physics, Beijing 100190, China }

\author{Wanxiang Feng}
\affiliation[Beijing Institute of Technology]
{School of Physics, Beijing Institute of Technology, Beijing 100081, China}

\author{Cheng-Cheng Liu}
\affiliation [Beijing Institute of Technology]
{School of Physics, Beijing Institute of Technology, Beijing 100081, China}

\author{Shan Guan}
\affiliation [Beijing Institute of Technology]
{School of Physics, Beijing Institute of Technology, Beijing 100081, China}

\author{Yugui Yao}
\email{ygyao@bit.edu.cn}
\affiliation [Beijing Institute of Technology]
{School of Physics, Beijing Institute of Technology, Beijing 100081, China}

\begin{document}

\begin{abstract}
Quantum spin Hall (QSH) insulators have gapless topological edge states inside the bulk band gap, which can serve as dissipationless spin current channels. The major challenge currently is to find suitable materials for this topological state.  Here, we predict a new large-gap QSH insulator with bulk direct band gap of $\sim$0.18 eV, in single-layer Bi$_{4}$Br$_{4}$, which could be exfoliated from its three-dimensional bulk material due to the weakly-bonded layered structure.  The band gap of single-layer Bi$_{4}$Br$_{4}$ is tunable via strain engineering, and the QSH phase is robust against external strain. Moreover, because this material consists of special one-dimensional molecular chain as its basic building block, the single layer Bi$_{4}$Br$_{4}$ could be torn to ribbons with clean and atomically sharp edges. These nano-ribbons, which have single-Dirac-cone edge states crossing the bulk band gap, are ideal wires for  dissipationless transport.  Our work thus provides a new promising material for experimental studies and practical applications of QSH effect. 


\begin{tocentry}
\includegraphics[width=8 cm]{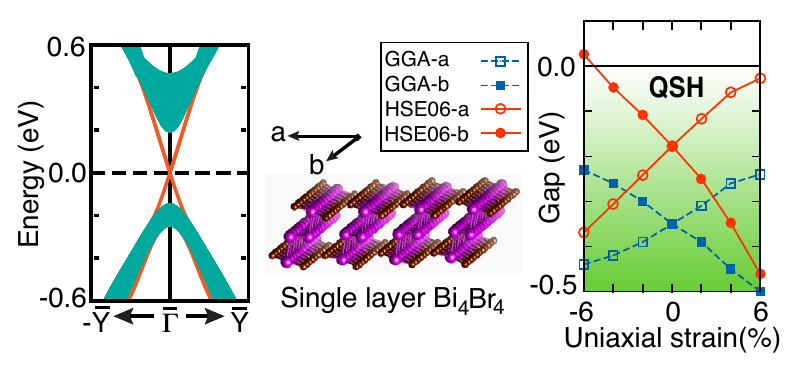}
\end{tocentry}

\end{abstract}

\newpage

Topological insulators (TIs) have generated a surge of interests recent years due to their rich physics and promising applications in spintronics and quantum computations~\cite{Hasan2010,Qi2011}. The concept of QSH insulators, also known as two-dimensional TIs, was first proposed in graphene~\cite{Kane2005}; however, later works demonstrated that the band gap opened by spin-orbit coupling (SOC) is extremely small ($10^{-3}$ meV) such that the QSH effect can not be observed in present experimental conditions~\cite{Yao2007,Huertas-Hernando2006,Min2006}.  Up to now, the QSH effect is only experimentally verified in HgTe/CdTe~\cite{Koenig2007} and InAs/GaSb~\cite{Knez2011} quantum wells, in which stringent conditions, \textit{e.g.} ultrahigh-quality samples and ultralow temperature, are required due to their small bulk band gaps (on the order of meV). Therefore, the search for QSH insulators with large band gaps and characteristics of easy fabrication is a key step towards the future studies and applications of QSH effect. 

Many efforts have been made to search for QSH insulators with large gap and stable structures,  and a number of  QSH insulators have been theoretically proposed, such as silicene~\cite{Liu2011a,*Liu2011b}, Bi(111) bilayer~\cite{Murakami2006}, bilayers of Group III elements with Bi~\cite{Chuang2014}, chemically modified Sn~\cite{Xu2013} and Bi/Sb~\cite{Song2014,*Liu2014} honeycomb lattices, and ZrTe$_{5}$/HfTe$_{5}$~\cite{Weng2014}. 
However, silicene has very small band gap, besides, an appropriate insulating substrate that can support the growth of silicene is still lacking; the Bi(111) bilayer can host sizable band gap, similar difficulty exists with the growth of the thin film material on a suitable substrate~\cite{Hirahara2011,Yang2012}; the chemically modified Sn, Sb and Bi honeycomb lattices are also difficult for experimental access, especially for the precise control of adatom coverage.
The most promising approach for fabricating QSH insulators may be by cleaving a chemically stable two-dimensional (2D) system from their van der Waals (vdW) layered three-dimensional (3D) matrix, just like graphene made by the scotch-tape method from graphite~\cite{Novoselov2004}.

In this Letter, based on first-principles calculations we predict that although the 3D bulk Bi$_{4}$Br$_{4}$ is a trivial insulator, its single layer form is a QSH insulator with large-gap of $\sim$0.18 eV.  The band gap can be effectively tuned by uniaxial strains, and the QSH phase is robust against external strains. The 3D bulk Bi$_{4}$Br$_{4}$ is a vdW layered semiconductor~\cite{Filatova2007}. Its interlayer binding energy is comparable to other layered systems that have been successfully exfoliated, such as graphite and MoS$_2$. Hence, the single layer Bi$_{4}$Br$_{4}$ could be made via the mechanical exfoliation from the bulk form. The phonon spectrum calculations further suggest that the freestanding single layer structure can be stable.  Moreover, the single layer Bi$_{4}$Br$_{4}$ has one-dimensional infinite molecular chain as its building block; therefore it could be naturally torn to nano-ribbons with clean and atomically sharp edges, which are much desired for the observation of topological edge states or serving as ideal one-dimensional (1D) dissipationless conducting wires. 

First-principles calculations are performed using the  projector augmented wave method~\cite{Blochl1994} implemented in the Vienna \textit{ab initio} simulation package~\cite{Kresse1993,*Kresse1996}. The generalized gradient approximation of  Perdew-Burke-Ernzerhof (GGA-PBE) is used for the exchange correlation potential~\cite{Perdew1996}. The structures are optimized employing the vdW corrections by the approach of Dion \textit{et al.}~\cite{klime2011,Dion2004}.  The experimental lattice parameters are used for the bulk system, and the same lattice parameters for single layer system (lattice constants a=13.064~\AA , b=4.338~\AA\ for Bi$_{4}$Br$_{4}$)\cite{Benda1978,Dikarev2001}.  
The ionic position are relaxed until force on each ion is less than 0.01 eV$\cdot$\AA$^{-1}$.  The phonon spectrum is calculated using the PHONOPY code~\cite{Togo2008} through the DFPT approach~\cite{Gonze1997} without SOC.
The Maximally Localized Wannier Functions (MLWFs) for the $p$ orbitals of Bi and Br atoms are constructed using the \textsc{wannier90} code\cite{Marzari1997,*Souza2001,Mostofi2008}.  Based on the generated MLWFs,  a tight-binding model for ribbons are constructed  to calculate the topological edge states.

\begin{figure}
\includegraphics[width=\columnwidth]{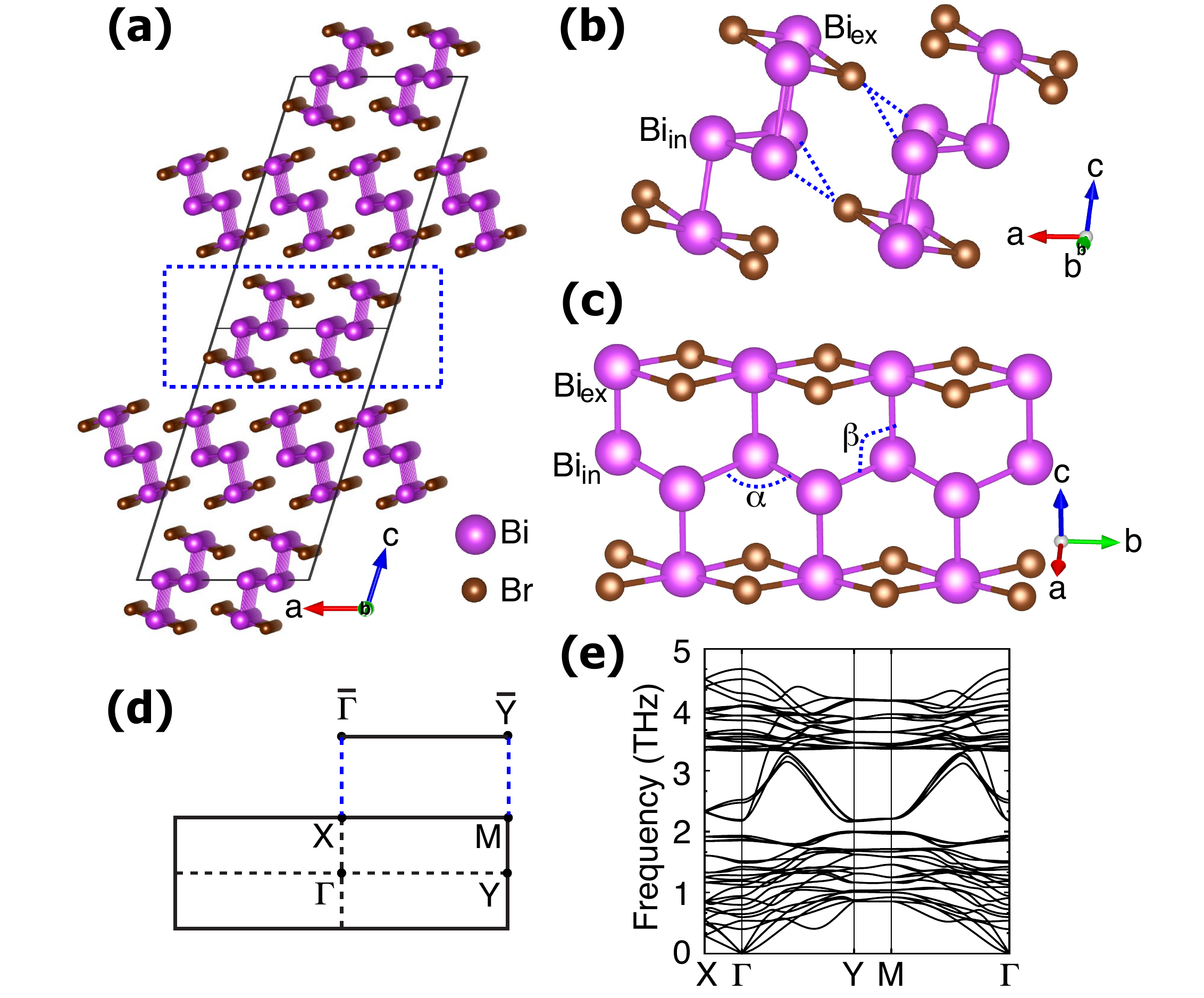}
\caption{Crystal structure of Bi$_{4}$Br$_{4}$ in 3D (a), 2D (b), and 1D (c) representations.  The large pink and small brown balls stand for Bi and Br atoms respectively. In the 1D chain, the Bi$_{in}$ atoms in the middle form a zigzag chain,  the Bi$_{ex}$ atoms at the edge of the chain are bonded with Br atoms.  (d) 2D and projected 1D Brillouin zones with high-symmetry points.  (e) phonon spectrum of single layer Bi$_{4}$Br$_{4}$. }\label{fig1}
\end{figure}

Bulk Bi$_{4}$Br$_{4}$ crystallizes in the monoclinic space group $C2/m$ ($C_{2h}^{3}$)~\cite{Benda1978,Dikarev2001}.\ref{fig1}a shows the layered structure of bulk Bi$_{4}$Br$_{4}$, which can be regarded as a combined packing of the normal (marked by blue dash line) and mirror-reflected single layer along the $c$-axis.  In each 2D layer, as shown in \ref{fig1}b, the basic building block is an 1D infinite molecular chain along the $b$-axis.  
\ref{fig1}c shows the structure of a single molecular chain. The middle Bi$_{in}$ atoms form a zigzag chain with strongly covalent Bi$_{in}$-Bi$_{in}$bonds ($\sim$3.0~\AA), and the Br atoms are tightly attached to Bi$_{ex}$ ($\sim$2.9~\AA) along the edges of the molecular chain. 
The Bi$_{in}$-Bi$_{in}$ and Bi$_{in}$-Bi$_{ex}$ bonds are nearly perpendicular to each other ($\alpha \sim 91^{\circ}$, $\beta \sim 92^{\circ}$). The bonds between Bi$_{ex}$ and Br atoms in the adjacent chains [dash line in \ref{fig1}b] are relatively weak ($\sim$3.6~\AA). 
Despite the weak bonding between adjacent chains,  the single-layer structure is rather stable,  given that all the synthesized mixed bismuth monohalides Bi$_4$Br$_{x}$I$_{4-x}$($x=1,2,3$) have the similar single-layer structure as Bi$_4$Br$_4$, while the inter-layer stacking patterns are diverse~\cite{Dikarev2001}.  

We have calculated the binding energy of the single-layer sheet to its bulk phase for Bi$_{4}$Br$_{4}$. 
The obtained value ($\sim$20 meV/\AA$^{2}$) is in the typical range for the vdW layered compounds~\cite{Bjorkman2012}. Specifically, the binding energy of single layer Bi$_{4}$Br$_{4}$ is comparable to the experimental result of graphene($\sim$ 12 meV/\AA$^{2}$)\cite{Liu2012} , and slightly smaller than theoretical value of MoS$_2$ ($\sim$26 meV/\AA$^{2}$) in our calculations.  
This indicates that the experimental fabrication of single layer Bi$_{4}$Br$_{4}$ become possible using similar ``scotch tape'' method as that for graphene~\cite{Novoselov2004,Geim2007} or MoS$_{2}$~\cite{Novoselov2005,Radisavljevic2011}. 
The dynamic stability of single layer Bi$_{4}$Br$_{4}$ is further investigated through the phonon spectrum calculations. The calculated phonon spectrum is shown in \ref{fig1}e. The phonon frequencies are real at all momenta, suggesting that the single layer structure of Bi$_{4}$Br$_{4}$ is dynamically stable.

\begin{figure*}
\includegraphics[width=16cm]{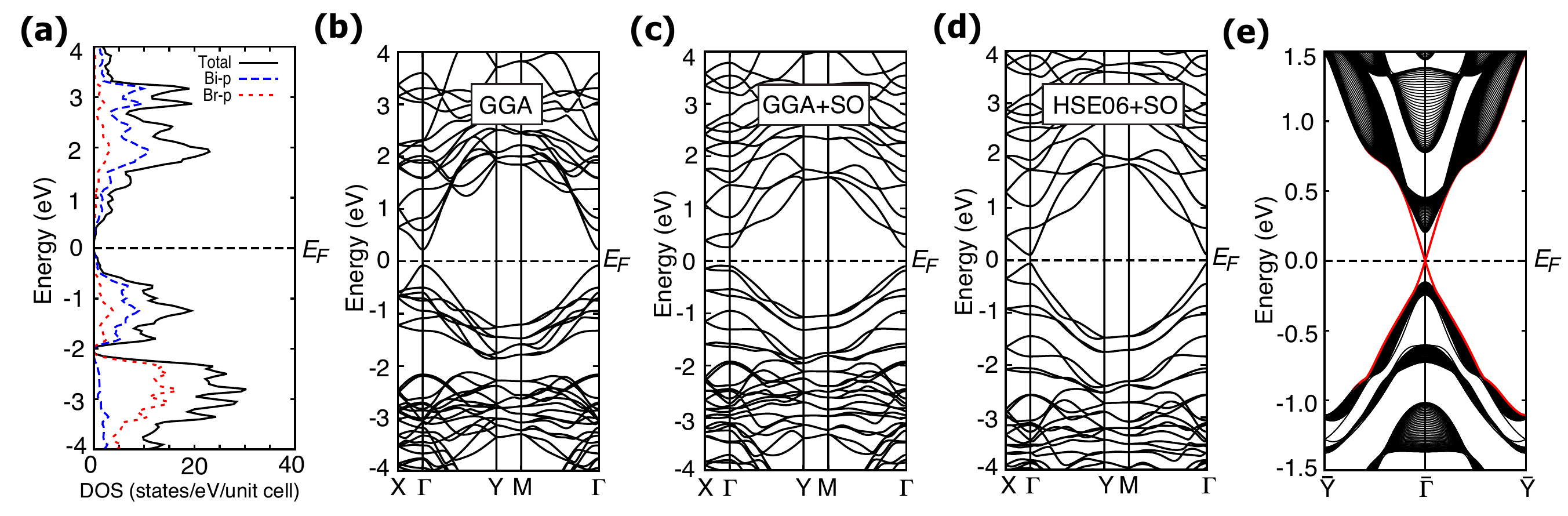}
\caption{
 Density of states (a) and  band structure  (b) of single-layer Bi$_{4}$Br$_{4}$ without SOC.  Band structures with SOC calculated by using (c) GGA and (d) HSE06 potentials. (e) Band structure of 1D nanoribbon edged along the molecular chain axis for Bi$_{4}$Br$_{4}$. Topological edge states are visualized by red lines crossing linearly at the $\Gamma$ point, which indicates single-layer Bi$_{4}$Br$_{4}$ is a QSH insulator. } \label{fig2}
\end{figure*}

We now focus on the topological properties of electronic structure of single-layer Bi$_{4}$Br$_{4}$.  
For the convenience of discussion, hereafter all of calculations are based on a 2D conventional cell that includes two 1D chain units [see \ref{fig1}b].  
The calculated band structure without SOC is plotted in \ref{fig2}b, in which one can see a direct band gap located at $\Gamma$ point. From the view of density of states shown in \ref{fig2}a, both of the valence and conduction bands near the Fermi level are mainly constituted by Bi-6$p$ orbitals, The states below -2 eV are dominated by Br-4$p$ orbitals due to the larger electronegativity of Br atoms.  
When SOC is switched on, the band gap maintains ($\sim$0.35 eV) as shown in \ref{fig2}c.  To analyze the band topology, we have calculated the Z$_{2}$ topological invariant within the first-principle framework~\cite{Feng2012}, based on the parity criterion proposed by Fu and Kane~\cite{Fu2007}.  The result of Z$_{2}$=1 verifies that the single layer Bi$_{4}$Br$_{4}$ is a QSH insulator.  
It is well known that the GGA-type calculation may produce inaccurate band gap. To further confirm the nontrivial properties of our system,  we have checked the band structure of Bi$_{4}$Br$_{4}$ using the more sophisticated Heyd-Scuseria-Ernzerhof hybrid functional method (HSE06)~\cite{Heyd2003}. The resultant band [see \ref{fig2}d] has a little smaller band gap ($\sim$0.18 eV) at $\Gamma$-point compared to the GGA result,  and the topological nontrivial phase remains unchanged.  

The gapless chiral edge states inside the bulk band gap is the hallmark of QSH insulators.  
Because the coupling between the 1D molecular chains is much weaker than the intra-chain bonding~\cite{Filatova2007}, a natural nanoribbon can be constructed along the $b$-axis without any dangling bond. With the nanoribbon width of 40 chains ($\approx$26 nm), the calculated band structure is presented in \ref{fig2}e. One can see explicitly that the gapless edge states appear in bulk band gap cross linearly at the $\Gamma$ point. The Fermi velocity of edge states in single layer Bi$_{4}$Br$_{4}$ is $\sim$4.8$\times 10^{5}$ m/s, which is slightly smaller than that of 5.5$\times 10^{5}$ m/s in HgTe/CdTe quantum well~\cite{Qi2011}, but is remarkably larger than that of 3$\times 10^{4}$ m/s in InAs/GaSb quantum well~\cite{Knez2011}.  This high Fermi velocity is beneficial to the fabrication of high-speed spintronics devices.  

\begin{figure}
\includegraphics[width=\columnwidth]{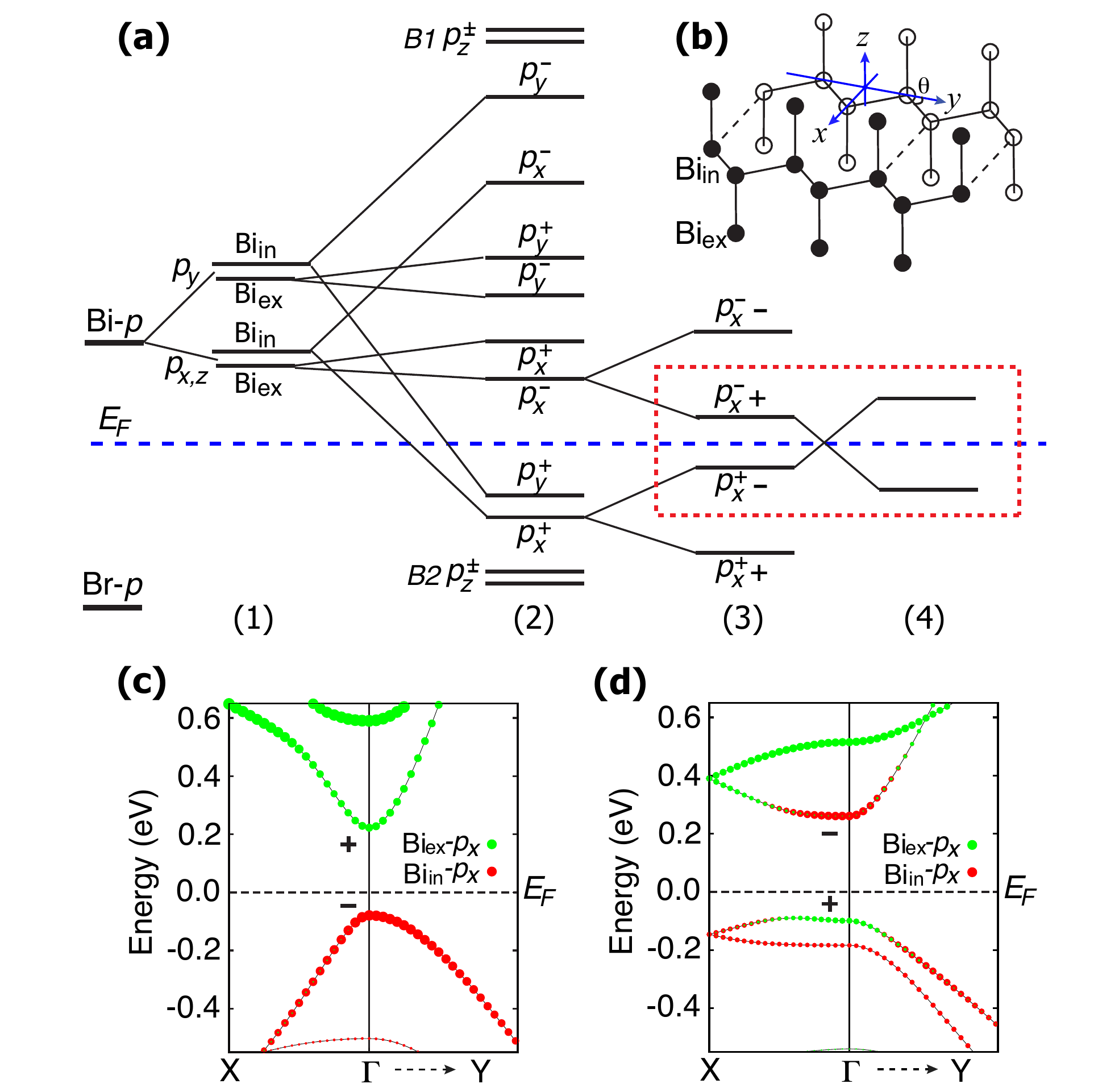}
\caption{
(a) Schematic diagram of the band evolution at $\Gamma$-point.  The evolution stages are explained in the main text. (b) Schematic figure of the single layer Bi$_{4}$Br$_{4}$ structure with Br-atoms neglected. The filled and empty circles stand for Bi-atoms from adjacent chains. 
The solid lines denote strong intra-chain bonds, the dash lines denote weak inter-chain coupling.  The Bi$_{in}$-$p_x$ and Bi$_{ex}$-$p_x$ orbital projected character of bands without (c) and with (d) SOC,  the parity is labeled for CBM and VBM at $\Gamma$-point.}
\label{fig3}
\end{figure}

To elucidate the physical mechanism of the band inversion,  we study the band structure evolution at $\Gamma$ point, starting from the atomic orbitals. The schematic diagram of the evolution is plotted in \ref{fig3}a.  The 2D single layer Bi$_4$Br$_4$ cell consists of two 1D chains,  and each chain has four Bi atoms and four Br atoms in a unit-cell. The Br-$4p$ orbitals stay deep below the Fermi level due to the much larger electronegativity of Bromine.  Therefore, it is reasonable to neglect the Br atoms, and only consider the Bi-$6p$ orbitals.  For clarity, the simplified structure with only Bi atoms is schematically plotted in \ref{fig3}b.

We first consider the strongly covalent bonds within the 1D chain,  and then take into account the weak inter-chain couping. For the 1D chain, the Bloch orbital at $\Gamma$ point is given by $|p_{\alpha}\rangle=\sum_{i}{p_{\alpha}^{i}}$, where $p_{\alpha}^{i}$ ($\alpha=x,y,z$) are the atomic orbitals in unit-cell $R_{i}$. At stage (1), mirror symmetry is considered.  
The structure has mirror symmetry with mirror plane parallel to $xz$ plane.  The $p_{x/z}$ and $p_y$ orbitals at $\Gamma$  point have different parities under mirror operation, hence they will not mix with each other (when SOC is absent).  The $p_y$ orbital has higher energy  because of the much stronger $\sigma$-type hopping between $p_{y}^{i}$ and $p_{y}^{i+1}$ compared to the $\pi$-type of $p_{x/z}$. At stage (2), the inversion symmetry of the chain is considered.  
Under inversion operation,  the two Bi$_{in}$ (Bi$_{ex}$) atoms interchange site with each other. Orbitals from these two atoms can be combined to form bonding and anti-bonding states with splitted levels, which are given by
\[
|Bi_{in/ex},\alpha^{\pm}\rangle =\frac{1}{\sqrt{2}}(|Bi_{in/ex}^{1},\alpha\rangle \mp|Bi_{in/ex}^{2},\alpha\rangle)
\]
with $\alpha=p_x, p_y, p_z$.   They have definite parities as denoted in the upper index $\pm$.
Only those states with the same parity can further couple with each other.
We discuss the intra-chain bonding induced coupling between those states as following:
(1)  The splitting energy $\Delta E_{x}$  between $|\textrm{Bi}_{in},p_{x}^{+}\rangle$ and $|\textrm{Bi}_{in},p_{x}^{-}\rangle$ is quite large due to the short Bi$_{in}$-Bi$_{in}$ bond length, so does the $|\textrm{Bi}_{in},p_{y}^{\pm}\rangle$. In addition, the relative size of $\Delta E_{x}$ and $\Delta E_{y}$ depends on the angle $\theta$ [see \ref{fig3}b].  
The $p_x$($p_y$) orbital hopping $t_{xx}$ ($t_{yy}$) between the two Bi$_{in}$ atoms, which determine the $\Delta E_{x}$ ($\Delta E_{y}$), is related to $\theta$,  namely,  $t_{xx}$ ($t_{yy}$) can be given by $t_{pp\sigma}{\textrm{sin}\theta}^2+t_{pp\pi}{\textrm{cos}\theta}^2$ ($t_{pp\sigma}{\textrm{cos}\theta}^2+t_{pp\pi}{\textrm{sin}\theta}^2$). 
Therefore,  the reduction of $\theta$ will decrease the $\Delta E_{x}$ and increase the $\Delta E_{y}$.
(2) The orbital hopping between the two Bi$_{ex}$ atoms, which locate at different sides of the chain, is negligible. However, $|\textrm{Bi}_{ex},p_{x/y}^{\pm}\rangle$ can be coupled to $|\textrm{Bi}_{in},p_{x/y}^{\pm}\rangle$ through the $\pi-$bonding of $p_{x/y}$ orbital between Bi$_{in}$-Bi$_{ex}$ atoms, which shifts $|\textrm{Bi}_{ex},p_{x/y}^{+}\rangle$  upward and $|\textrm{Bi}_{ex},p_{x/y}^{-}\rangle$ downward.
(3) Due to the strong $\sigma-$bonding of $p_z$ orbital between Bi$_{in}$-Bi$_{ex}$ atoms,  $|\textrm{Bi}_{ex},p_{z}^{\pm}\rangle$ mix with $|\textrm{Bi}_{in},p_{z}^{\pm}\rangle$ and splits into $|\textrm{B}1, p_{z}^{\pm}\rangle$ and $|\textrm{B}2, p_{z}^{\pm}\rangle$ states. 

So far, we have considered all the intra-chain bonding.  At stage (3), we start to count in the inter-chain coupling.  The inter-chain coupling is dominated by the hopping between $p_x$ orbitals because their orientations are parallel to the inter-chain direction.  
The $p_{x}^{\pm}$ orbitals from two adjacent chains (labeled as A, B) within the 2D cell are further splitted into bonding and anti-bonding states, for example
\[
|Bi_{in/ex},p_{x}^{+}, \pm\rangle =\frac{1}{\sqrt{2}}(|Bi_{in/ex}^{A},p_{x}^{+}\rangle \mp|Bi_{in/ex}^{B},p_{x}^{+}\rangle)
\]
At final stage (4), the SOC is considered. The SOC mixes the orbital and spin with total angular momentum conserved, which results in the level repulsion between $|\textrm{Bi}_{in/ex}, p_{x}^{\pm}\rangle$ and $|\textrm{Bi}_{in/ex}, p_{y}^{\pm}\rangle$ \cite{Liu2010}, thus push the $|\textrm{Bi}_{ex}, p_{x}^{-}, +\rangle$ downward and the $|\textrm{Bi}_{in}, p_{x}^{+}, -\rangle$ upward, as shown in \ref{fig3}a. Consequently, the band order is inverted (marked by the red dashed box).
We show the Bi$_{in}$-$p_x$ and Bi$_{ex}$-$p_x$ orbitals projected bands in \ref{fig3}c (without SOC) and \ref{fig3}d (with SOC). As we can seen, in the absence of SOC,  the valence band maximum(VBM) is dominated by the Bi$_{in}$-$p_x$ orbital with negative parity,  while the conduction band minimum (CBM) is dominated by the Bi$_{ex}$-$p_x$ orbital with positive parity.  After SOC is turned on, both the orbital character and the parity of CBM and VBM are inverted.

\begin{figure}
\includegraphics[width=\columnwidth]{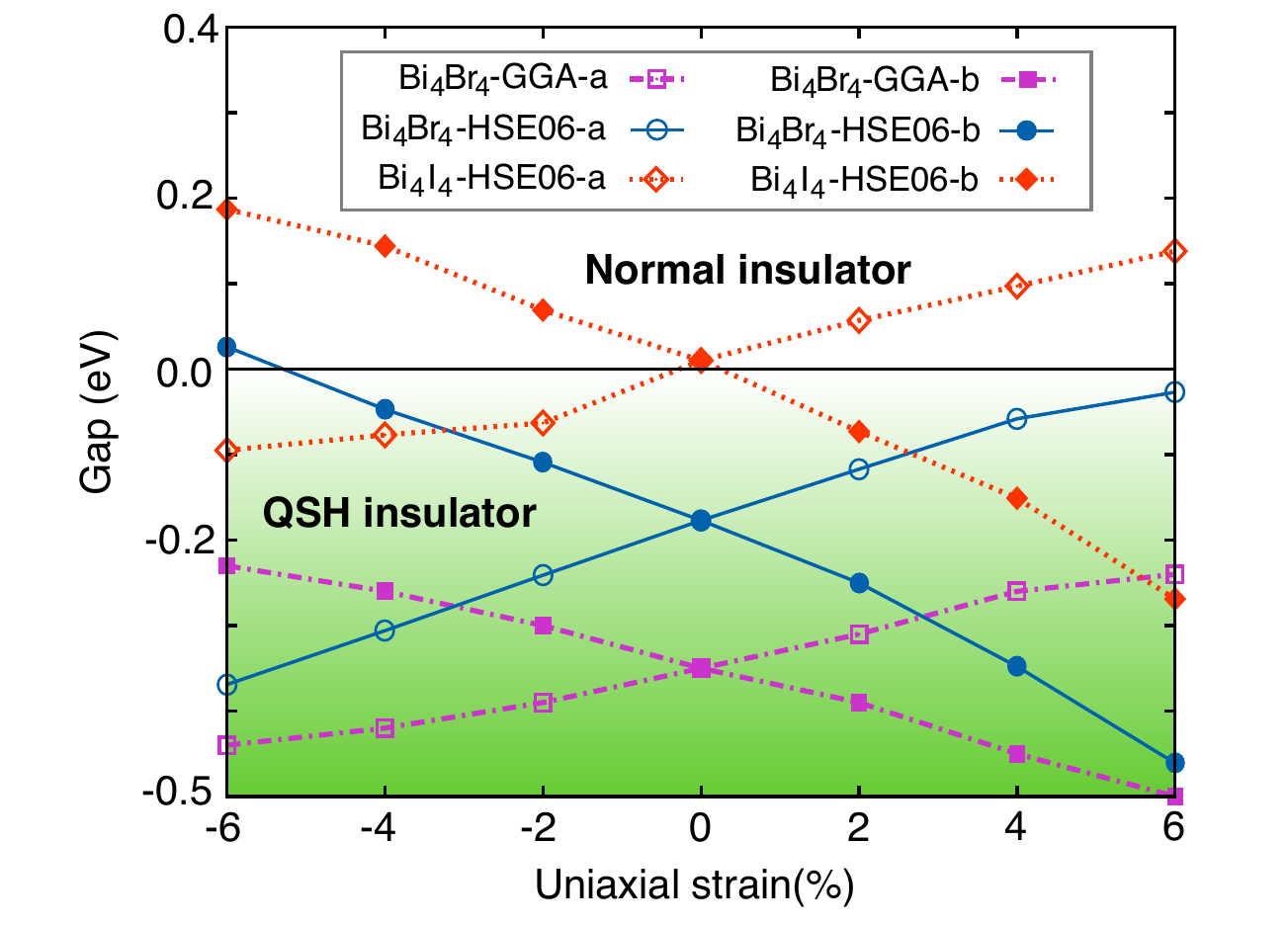}
\caption{Band gaps as a function of uniaxial strains. The negative band gaps indicate non-trivial topological phases. The uniaxial strains along a- and b-axises are adopted separately. Band gaps of single layer Bi$_{4}$Br$_{4}$ calculated by using GGA-PBE and HSE06 potentials, and band gaps of single layer Bi$_{4}$I$_{4}$ calculated by using HSE06 potential are presented respectively. }\label{fig4}
\end{figure}

Application of external strain can effectively modify the band gap of single layer Bi$_{4}$Br$_{4}$, as shown in \ref{fig4}.  Due to the structural anisotropy, we adopt separately the uniaxial strains along $a$- and $b$-axises.  
The change of band gap under the uniaxial strains displays a opposite behavior, i.e. increasing the lattice constant $a$ will diminish the band gap, while increasing the lattice constant $b$ will enlarge the band gap. 
This behavior can be understood from the above band evolution analysis. Increasing $a$ will extend the inter-chain distance, which decreases the inter-chain coupling. The weakened inter-chain coupling will diminish the splitting at stage (3) in \ref{fig3}a and decrease the band gap ultimately.  As we increase the lattice constant $b$,  the angle $\theta$ becomes smaller accordingly while the Bi$_{in}$-Bi$_{in}$ bond length hardly changes.  With smaller $\theta$, the $|\textrm{Bi}_{in},p_{x}^{+}\rangle$ will move upward to reduce the splitting [stage (2) in \ref{fig3}a], which then enlarges the band gap ultimately.  

Although the band gap is tunable via strain engineering, the topological nontrivial phase of single layer Bi$_{4}$Br$_{4}$ survives in a large range of strains from $\sim$ -6\% to 6\%. Hence, the QSH phase can be robust against lattice mismatch strain when the single layer Bi$_{4}$Br$_{4}$ is supported by a substrate. We have also investigated another bismuth monohalides Bi$_{4}$I$_{4}$, which has the similar structure as Bi$_{4}$Br$_{4}$~\cite{Benda1978,Dikarev2001}. The single layer Bi$_{4}$I$_{4}$ is near the boundary of topological trivial and nontrivial phase transition[see \ref{fig4}]; hence its QSH phase can be effectively tuned by uniaxial strains. 

Finally, we extend our discussion to the multilayer structures of Bi$_{4}$Br$_{4}$.  
Usually, when the single-layers are stacked together to form a multilayer film or 3D compound, the band gap will be reduced due to the inter-layer orbital hopping. 
In our systems, the inter-layer coupling is rather weak and the band edges are all dominated by the in-plane orbitals; therefore, we do not expect much band gap reduction in multilayer Bi$_{4}$Br$_{4}$ compared to the single layer one. Even for the 3D compound, the inverted band gap ($\sim$0.15 eV) is only reduced by $\sim$30 meV according to our HSE06 calculations. Within the weak coupling limit, the multilayer Bi$_{4}$Br$_{4}$ with even (odd) number of layers would have even (odd) times of band inversions at $\Gamma$ points~\cite{Yan2012}. Therefore, the $Z_2$ topological invariant would exhibit an interesting even-odd oscillation with increasing number of layers. The 3D bulk Bi$_{4}$Br$_{4}$, which contains two layers in one unit cell, is a trivial insulator, while the ultra-thin films with odd number of layers are QSH insulators~\cite{Yan2012}. 

In summary, the 2D Bi$_{4}$Br$_{4}$ is a promising QSH insulator with large band gap ($\sim$0.18 eV), while the single layer Bi$_{4}$I$_{4}$ is near the boundary of topological trivial and nontrivial phase transition. Due to the highly structural anisotropy,  natural Bi$_{4}$Br$_{4}$ nonaribbons can be constructed without any dangling bond.  These nanoribbons, which have single-Dirac-cone topological edge states inside the bulk band gap,  can serve as ideal 1D wires for dissipationless transport, or be utilized to create and manipulate majorana fermion via, e.g.  putting the nanoribbon on the s-wave superconductors with external magnetic fields~\cite{Alicea2012}. 

\acknowledgement
This work was supported by the MOST Project of China (Nos.~2014CB920903, 2013CB921903, 2011CBA00100), the NSF of China (Nos.~11174337, 11225418, 11374033) and the SRFDPHE of China (No.~20121101110046, 20131101120052).

\bibliography{bix-ref}

\end{document}